\documentclass[10pt,twocolumn,showpacs,pre,amsmath,amssymb,floatfix]{revtex4-1}
\usepackage{amsfonts}
\usepackage{amsmath,amssymb,color}
\usepackage{bm}
\usepackage{graphicx}
\usepackage{psfrag}
\usepackage{ulem}
\newcommand{\bd}{\bm}

\begin{document}

\title{Renormalization group for the $\varphi^4$-theory with long-range interaction \\ and the
critical exponent $\eta$ of the Ising model}

\author{Raphael Goll}
\affiliation{Institut f\"{u}r Theoretische Physik, Universit\"{a}t Frankfurt,  Max-von-Laue Stra{\ss}e 1, 60438 Frankfurt, Germany}

\author{Peter Kopietz}
\affiliation{Institut f\"{u}r Theoretische Physik, Universit\"{a}t Frankfurt,  Max-von-Laue Stra{\ss}e 1, 60438 Frankfurt, Germany}

\date{September 17, 2018}

 \begin{abstract}
We calculate the  critical exponent $\eta$ of the $D$-dimensional Ising model from a simple truncation of the functional renormalization group flow equations for  a scalar field theory with long-range interaction.
Our approach relies on the smallness of the inverse range of the interaction and on the assumption  that the Ginzburg momentum defining the width of the scaling regime in momentum space is larger than the scale where the renormalized interaction crosses over from long-range to short-range;
the numerical value of $\eta$ can then be estimated by stopping the renormalization group  flow at this scale.
In three dimensions our result $\eta = 0.03651$ is in good agreement with recent conformal bootstrap and Monte Carlo calculations.
We extend our calculations to fractional dimensions $D$ and obtain the resulting critical exponent $\eta(D)$ between two and four dimensions.
For dimensions $2\leq D \leq 3$ our result for $\eta$ is consistent with previous calculations.
\end{abstract}

\maketitle

\section{Introduction}
\vspace*{-.3cm}
The calculation of the precise numerical values of the critical exponents which characterize the power-law singularities of various thermodynamic observables in the vicinity of continuous phase transitions remains one of the big challenges in theoretical physics. 
Although the renormalization group (RG) theory developed by Wilson, Fisher, and others \cite{Wilson71,Wilson72a,Wilson72b,Fisher74,Wilson74,Wegner76,DiCastro76,Ma76} in the 1970s provides a deep understanding of the origin of the universality of the critical exponents, 
a controlled calculation of their numerical values for systems whose dimensionality $D$ lies below  the so-called upper critical dimension $D_u$ is very difficult due to the absence of a small parameter.
Successful strategies to solve this problem for the Ising universality class in three dimensions included an expansion of the critical exponents in powers of $\epsilon = D_u - D$ (and careful extrapolation to $D=3$) \cite{LeGuillou77,LeGuillou87,Yukalov98}, 
fixed dimension expansion methods \cite{Guida98,Kleinert01}, and high-temperature series expansions \cite{Campostrini02}.
Another class of precise estimates for the critical exponents was obtained by a variety of numerical methods, such as Monte Carlo simulations in combination with finite size scaling analysis \cite{Bloete99,Hasenbusch10}, extensions thereof taking into account cross correlation \cite{Weigel09}, or Monte Carlo renormalization group approaches \cite{Bloete96,Ron17}.
The most precise estimates to date are provided by the conformal bootstrap method \cite{El-Showk12,Kos16}
and a recent Monte Carlo study \cite{Ferrenberg18}, fixing e.g the value  of the anomalous dimension to $\eta=0.0362978(20)$.
Recent sophisticated truncations of the exact functional renormalization group (FRG) flow equation for the average effective action \cite{Wetterich93} based on either the derivative expansion \cite{Berges02,Canet03} or the so-called Blaizot-M\'endez-Wscheebor (BMW) approximation \cite{Benitez12} have produced results for $\eta$ which lie within 10 $\%$ from the above mentioned values.

In this work we develop a new and remarkably simple method for calculating the exponent $\eta$ of the Ising universality class in arbitrary dimensions. 
Our method is based on a simple truncation of the exact FRG flow equations of a scalar field theory with long-range quartic interaction. 
We show that the inverse range of the interaction can be used as a small parameter to control the truncation of the hierarchy of FRG flow equations. 
Although the range of the renormalized interaction decreases as the RG is iterated, we show that by stopping the RG flow at some finite scale (which will be uniquely defined below) we obtain the critical exponent $\eta$ of the Ising model in $D=3$ ($D=2$) within a precision of about $10 \%$ ($20 \%$).  
The fact that in $D=3$ our result $\eta = 0.03651$ deviates significantly less from the accepted results \cite{Ferrenberg18,Kos16} indicates  that in this case our method may actually be more precise than anticipated.

\vspace*{-.35cm}
\section{Long-Range $\phi^4$ Model}
\vspace*{-.3cm}
Our starting point is the following action for a real scalar field $\varphi ( \bd{x} )$ in $D$ dimensions with long-range interaction $V_0 ( \bd{x} )$,
 \begin{eqnarray}
 {S} [ \varphi ] &  =  &  \frac{1}{2} \int d^D x \Bigl[
 r_{0} \varphi^2 ( \bd{x} )
 +
 c_{0} ( \bd{\nabla} \varphi ( \bd{x}) )^2  
 \Bigr]
 \nonumber
 \\
 & + &  \frac{1}{8} \int d^D x \int d^D x^{\prime}
 \varphi^2 ( \bd{x} ) V_0 ( \bd{x} - \bd{x}^\prime ) \varphi^2 ( \bd{x}^{\prime} ),
 \hspace{7mm}
 \label{eq:Sbare}
 \end{eqnarray}
where  $r_0$ is proportional to the inverse order-parameter susceptibility in mean-field approximation, 
the constant $c_0$ is  positive, 
and a short-distance cutoff $1/ \Lambda_0$ of the order of the lattice spacing of the underlying Ising model is implicit.
We assume that the  Fourier transform $V_0 ( \bd{k} )$ of the interaction is for wavevectors $\bd{k} \neq 0$ given by
 \begin{equation}
 V_0 ({\bd{k}}) =  \int d^{D} x e^{ - i \bd{k} \cdot \bd{x}} V_0 ( \bd{x} ) =
 \frac{ 1}{ m_0 + b_0 \bd{k}^2} ,
 \label{eq:F0def}
 \end{equation}
with positive constants $m_0$ and $b_0$.
Moreover, precisely for $\bd{k} =0$ we set $V_0 ( \bd{k}=0) =0$, 
so that the perturbative expansion does not contain tadpole diagrams such as the Hartree contribution to the self-energy, analogously to electronic systems where this is required by charge neutrality \cite{FetterWalecka,Amit05}.
In three dimensions  Eq.~(\ref{eq:F0def}) corresponds to the screened Coulomb interaction 
$V_0 ( \bd{x} ) = e^{ - \kappa_0 | \bd{x} | }/(4 \pi b_0 | \bd{x} | )$, 
where the wavevector 
$\kappa_0 = \sqrt{ m_0 / b_0 }$ 
can be identified with the inverse range of the interaction.
We assume that $\kappa_0 $ is much smaller than the ultraviolet cutoff $\Lambda_0$.
The small parameter
$ \kappa_0 / \Lambda_0 $ 
will play an important role for controlling the precision of our truncation of the FRG flow equations.
Note that in the usual Ginzburg-Landau-Wilson functional \cite{Ma76,Kopietz10} describing the long-wavelength order-parameter fluctuations of the Ising model the interaction is usually assumed to be local, 
$V_0 ( \bd{x} ) \propto \delta ( \bd{x} )$.
However, because of universality, 
we can also obtain the critical exponents of the Ising universality class from the long-range interaction model~(\ref{eq:F0def}).

Due to the long-range nature of the interaction, our action (\ref{eq:Sbare}) is non-local, therefore approximation strategies based on the local potential approximation \cite{Canet03} cannot be used.
However, we can make our action (\ref{eq:Sbare}) local by decoupling the interaction by means of a real Hubbard-Stratonovich field $\psi ( \bd{x} )$ conjugate to the composite field
$\varphi^2 ( \bd{x} )$.
In momentum space the decoupled action is then
\begin{eqnarray}
 {S} [ \varphi , \psi ]  & = &  
 \frac{1}{2} \int_{\bd{k}}
 \bigl[
  G_0^{-1} ( \bd{k} ) \varphi_{ - \bd{k} }  \varphi_{ \bd{k} } 
 +   V_0^{-1} ( {\bd{k}} )  \psi_{ - \bd{k} } \psi_{ \bd{k} }
 \bigr]
 \nonumber
 \\
 &  + &  \frac{i}{2} \int_{\bd{k}_1} \int_{\bd{k}_2} \int_{\bd{k}_3}
 \delta_{ \bd{k}_1 + \bd{k}_2 + \bd{k}_3 ,0}
\psi_{  \bd{k}_1 } \varphi_{  \bd{k}_2 }
 \varphi_{  \bd{k}_3 },
 \label{eq:Shs}
 \end{eqnarray}
with 
$G_0^{-1} ( \bd{k} )  =   r_0 + c_0 {k}^2$ 
and 
$V_0 ( \bd{k} )$ 
given in Eq.~(\ref{eq:F0def}). 
We have introduced the notation 
$\int_{\bd{k}} = \int d^D k /(2 \pi )^D$
and 
$\delta_{ \bd{k} , 0 } = (2 \pi )^D \delta ( \bd{k} )$.
Because the fields $\varphi ( \bd{x} )$ and $\psi ( \bd{x} )$ are real, 
their  Fourier components satisfy
$\varphi_{ - \bd{k} } = \varphi^{\ast}_{\bd{k} }$ 
and
$\psi_{ - \bd{q} } = \psi^{\ast}_{ \bd{q} }$.
Note however, that the conjugate field $\psi$ does not represent a relevant physical excitation but rather a mediator of the long range interaction.

\begin{figure}[t]
 \begin{center}
  \centering
\vspace{3mm}
 \includegraphics[width=0.45\textwidth]{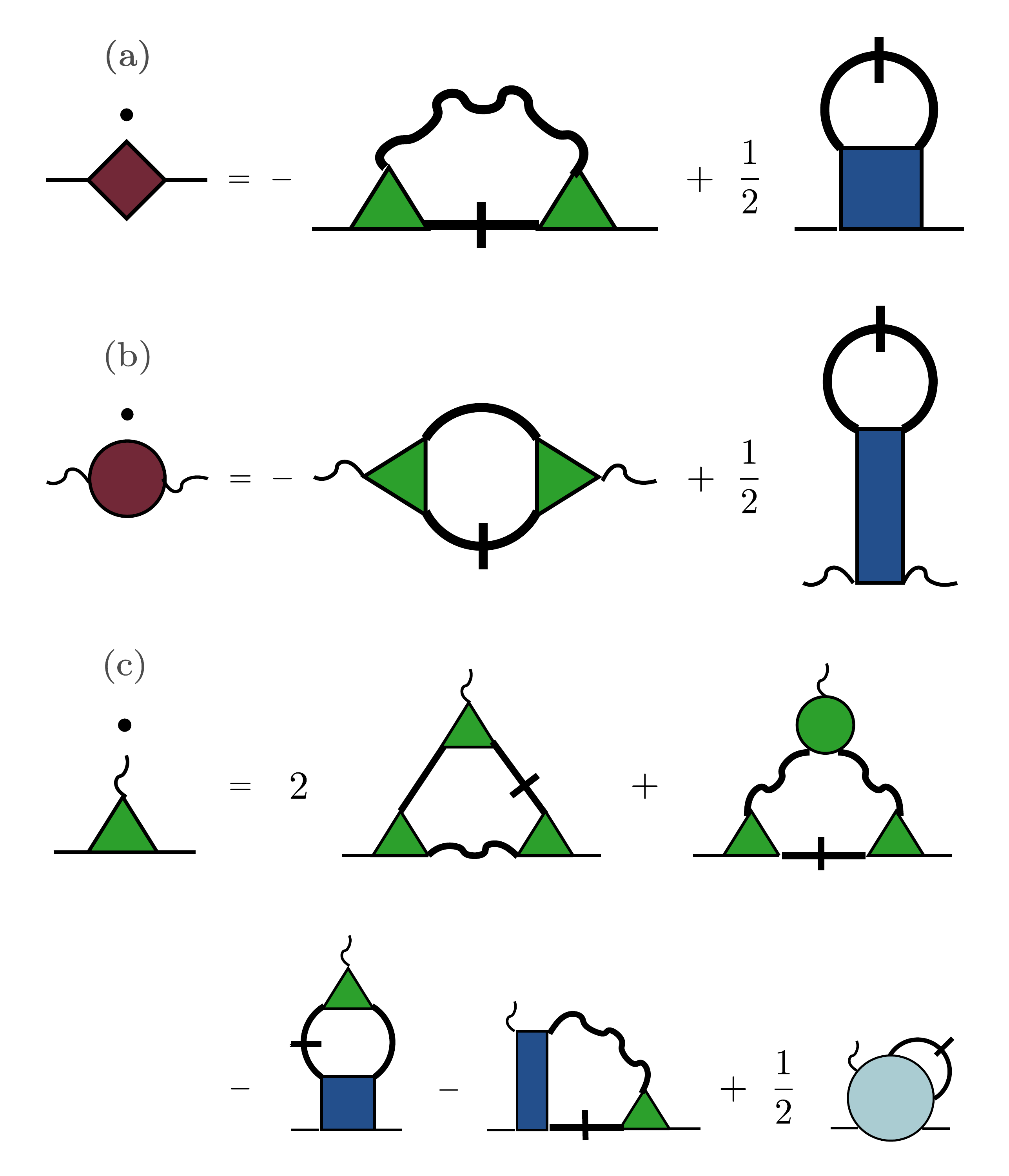}
   \end{center}
  \caption{
Graphical representations of exact FRG flow equations:
(a)  Self-energy $\Sigma_{\Lambda} ( \bd{k} )$ of the
order-parameter field $\varphi$;  (b) 
self-energy $\Pi_{\Lambda} ( \bd{k} )$ of
the conjugate field $\psi$;
(c) mixed three-legged vertex
$\Gamma_{\Lambda}^{ \psi \varphi \varphi } ( \bd{k}_1 ; \bd{k}_2 , \bd{k}_3)$.
 Here the solid lines represent the
cutoff-dependent $\varphi$-propagator $G_{\Lambda} ( \bd{k} )$,
while the wiggly lines represent the
$\psi$-propagator $V_{\Lambda} ( \bd{k} )$.
Slashed lines represent the corresponding single-scale propagators of the order parameter field.
}
\label{fig:flowself}
\end{figure}

\vspace*{-.35cm}
\section{FRG Approach}
\vspace*{-.3cm}
Following the usual procedure we now replace the propagator of the order parameter field by the cutoff-dependent deformation $G_0 ( \bd{k} ) \rightarrow G_{ 0 , \Lambda} ( \bd{k} )$ such that for $ k \lesssim \Lambda$ the deformed propagator is small while for $k \gtrsim \Lambda$ we recover the bare propagator.
Since we are eventually interested in the behavior close to the Wilson-Fisher fixed point the cutoff is thereby introduced only in the order parameter field.
It is then straightforward \cite{Kopietz10} to write down formally exact FRG flow equations for the irreducible vertices of the model (\ref{eq:Shs}) describing the evolution of these vertices when the cutoff parameter $\Lambda$ is reduced. 
Of particular interest are the flow equations of the self-energy
$\Sigma_{\Lambda} ( \bd{k} )$ of the order-parameter field $\varphi$ 
and the self-energy $\Pi_{\Lambda} ( \bd{k} )$ of the conjugate field $\psi$, 
which are shown graphically in Fig.~\ref{fig:flowself}.
Obviously, these flow equations depend on various higher order irreducible vertices with three and four external legs. 
For our purpose it is fortunately sufficient to retain only the mixed three-legged vertex
$\Gamma^{ \psi \varphi \varphi }_{\Lambda}  
( \bd{k}_1 ; \bd{k}_2 , \bd{k}_3)$
which is the only higher-order vertex with a  non-zero initial value
$\Gamma^{ \psi \varphi \varphi }_{\Lambda_0}  
( \bd{k}_1 ; \bd{k}_2 , \bd{k}_3) = i$.
The exact flow equation for this vertex is shown graphically in Fig.~\ref{fig:flowself} (c).
Neglecting all other vertices on the right-hand sides of the flow equations in Fig.~\ref{fig:flowself}, we obtain a closed system of integro-differential equations for the three functions 
$\Sigma_{\Lambda} ( \bd{k} )$, $\Pi_{\Lambda} ( \bd{k} )$ and
$\Gamma^{ \psi \varphi \varphi }_{\Lambda}  
( \bd{k}_1 ; \bd{k}_2 , \bd{k}_3)$.
To further reduce the complexity of the problem, we neglect the momentum-dependence of the three-legged vertex, setting
$\Gamma^{ \psi \varphi \varphi }_{\Lambda}  
( \bd{k}_1 ; \bd{k}_2 , \bd{k}_3) \approx i \gamma_{\Lambda}$.
As discussed below, this approximation is controlled as long as the renormalized interaction is long-range, and we then obtain the closed system of integro-differential equations,
 \begin{eqnarray}
  \partial_{\Lambda} \Sigma_{\Lambda} ( \bd{k} )  & =  &
 \gamma_{\Lambda}^2     \int_{\bd{q}} 
\dot{G}_\Lambda ( \bd{q} )   {V}_{\Lambda} ( \bd{q} + \bd{k} ),
 \label{eq:flowSigma3}
 \\
  \partial_{\Lambda} \Pi_{\Lambda} ( \bd{k} )  & =  &
   \gamma_{\Lambda}^2 \int_{\bd{q}}  
 \dot{G}_{\Lambda} ( \bd{q} ) G_{\Lambda} ( \bd{q} + \bd{k} ) ,
 \label{eq:flowPi3}
 \\
 \partial_{\Lambda} \gamma_{\Lambda}  & = & - 2 \gamma_{\Lambda}^3 \int_{\bd{q}}
  \dot{G}_{\Lambda} ( \bd{q} ) G_{\Lambda} ( \bd{q} ) V_{\Lambda} ( \bd{q} ).
 \label{eq:gammaflowtrunc}
 \end{eqnarray}
Choosing a sharp momentum cutoff, the propagator and single-scale propagator of the order-parameter field are
  \begin{eqnarray}
  G_{ \Lambda} ( \bd{k} ) & = & \frac{ \Theta ( k - \Lambda )}{ 
 r_0 + c_0 k^2 + \Sigma_{\Lambda} ( \bd{k} ) },
 \\
 \dot{G}_{ \Lambda} ( \bd{k} ) & = & - \frac{ \delta ( k - \Lambda )}{ 
 r_0 + c_0 k^2 + \Sigma_{\Lambda} ( \bd{k} ) },
 \end{eqnarray}
while the scale-dependent effective interaction is
 \begin{align}
V_\Lambda(\bd{k})=
\frac{1}{m_0+b_0 k^2+\Pi_\Lambda(\bd{k})}.
\end{align}
Equations.~(\ref{eq:flowSigma3})--(\ref{eq:gammaflowtrunc}) form a closed system of integro-differential equations for the two self-energies $\Sigma_\Lambda ( \bd{k} )$ and $\Pi_{\Lambda} ( \bd{k})$ and the vertex $\gamma_{\Lambda}$. 
Although these equations may in principle be solved numerically without further approximations,
to make progress analytically we expand the self-energies for small momenta up to order $k^2$,
 \begin{eqnarray}
 \Sigma_{\Lambda} ( \bd{k} ) & = & r_{\Lambda} - r_0  + ( c_{\Lambda}  - c_0 ) k^2 +
 {\cal{O}} ( k^4 ),
 \label{eq:sigmalong}
 \\
 \Pi_\Lambda ( \bd{k} ) & =& m_{\Lambda} - m_0  + a_{\Lambda} | \bd{k} |    + 
 ( b_\Lambda - b_0 ) k^2      + {\cal{O}} ( k^4 ).
 \hspace{7mm}
 \label{eq:pilong}
 \end{eqnarray}
Note that for sharp momentum cutoff the expansion of 
$\Pi_{\Lambda} ( \bd{k} )$
has a non-analytic term proportional to $|  \bd{k} |$~\cite{Ledowski04}.
Substituting the expansions (\ref{eq:sigmalong}) and (\ref{eq:pilong}) into our flow equations (\ref{eq:flowSigma3})--(\ref{eq:gammaflowtrunc}) it is straightforward to derive RG flow equations for the six couplings 
$r_{\Lambda}$, $c_{\Lambda}$, $m_\Lambda$, $a_{\Lambda}$, $b_{\Lambda}$ and $\gamma_\Lambda$. 
In order to find the scaling solution corresponding to the Wilson-Fisher fixed point it is convenient to introduce the dimensionless rescaled couplings 
$r_l = r_{\Lambda} / ( c_{\Lambda} \Lambda^2)$,
$a_l = a_{\Lambda} \Lambda / m_{\Lambda}$, and $b_l = b_{\Lambda} \Lambda^2 / m_{\Lambda}$, 
which are considered as functions of the logarithmic flow parameter
$l = \ln ( \Lambda_0 / \Lambda )$.
This choice furthermore reveals that the two parameters $m_\Lambda$ and $\gamma^2_\Lambda$ only appear in the combination $m_\Lambda/\gamma^2_\Lambda$ in the resulting flow equations. 
It is thus natural to reduce the number of relevant parameters by one by introducing the rescaled version of the momentum independent part of the interaction as 
$g_l = \Omega_D \Lambda^{D-4}  \gamma^2_{\Lambda} /\big( (2 \pi)^D c _{\Lambda}^2 m_{\Lambda} \big) $,
where $\Omega_D $ is the surface area of the $D$-dimensional unit sphere.
We obtain 
\begin{eqnarray}
  \partial_l r_l & = & ( 2 - \eta_l ) r_l + \frac{g_l}{(1 + r_l)s_l } ,
 \label{eq:rlflow}
 \\
 \partial_l g_l & = &  ( 4 -D - 2 \eta_l ) g_l - \frac{ g_l^2}{ ( 1 + r_l )^2 }
 \left[ \frac{1}{2} + \frac{2}{s_l}  \right],
 \label{eq:ulflow}
 \\
 \partial_l a_l & = & - \left[ 1 + \frac{g_l}{ 2 ( 1 + r_l )^2 } \right] a_l
-   \frac{g_l}{2} \frac{ \Omega_{D-1}}{ \Omega_D}
   \frac{ \frac{5-D}{D-1}  -   r_l  }{  ( 1 + r_l )^3 },
  \label{eq:alflow}
 \hspace{7mm}
 \\
 \partial_l b_l & = & - \left[ 2 + \frac{g_l}{ 2 ( 1 + r_l )^2 } \right] b_l
 +  \frac{g_l}{2} \frac{  \frac{ 4-D}{D}  -  r_l }{ ( 1 + r_l )^4 },
 \label{eq:bflow}
 \hspace{7mm}
 \end{eqnarray}
where the scale-dependent coupling
$s_l = 1 + a_l + b_l$ is large if the interaction is long-range and
reduces to a number  of order unity for short-range interaction.
The flowing anomalous dimension $\eta_l = - (\Lambda \partial_{\Lambda} c_{\Lambda}) / c_{\Lambda}$  is given by
 \begin{eqnarray}
  \eta_l  &= &  
 \frac{g_l    }{ ( 1+ r_l ) s_l^2}  \left[ \frac{ ( a_l + 2 b_l )^2}{D s_l} - \frac{ D-1}{2D} a_l -  b_l
 \right].
 \label{eq:etal}
  \end{eqnarray}
In the rest of this work, we carefully analyze the RG flow encoded in
Eqs.~(\ref{eq:rlflow})--(\ref{eq:etal}).

 \begin{figure}[b]
 \begin{center}
  \centering
\vspace{3mm}
 \includegraphics[width=0.475\textwidth]{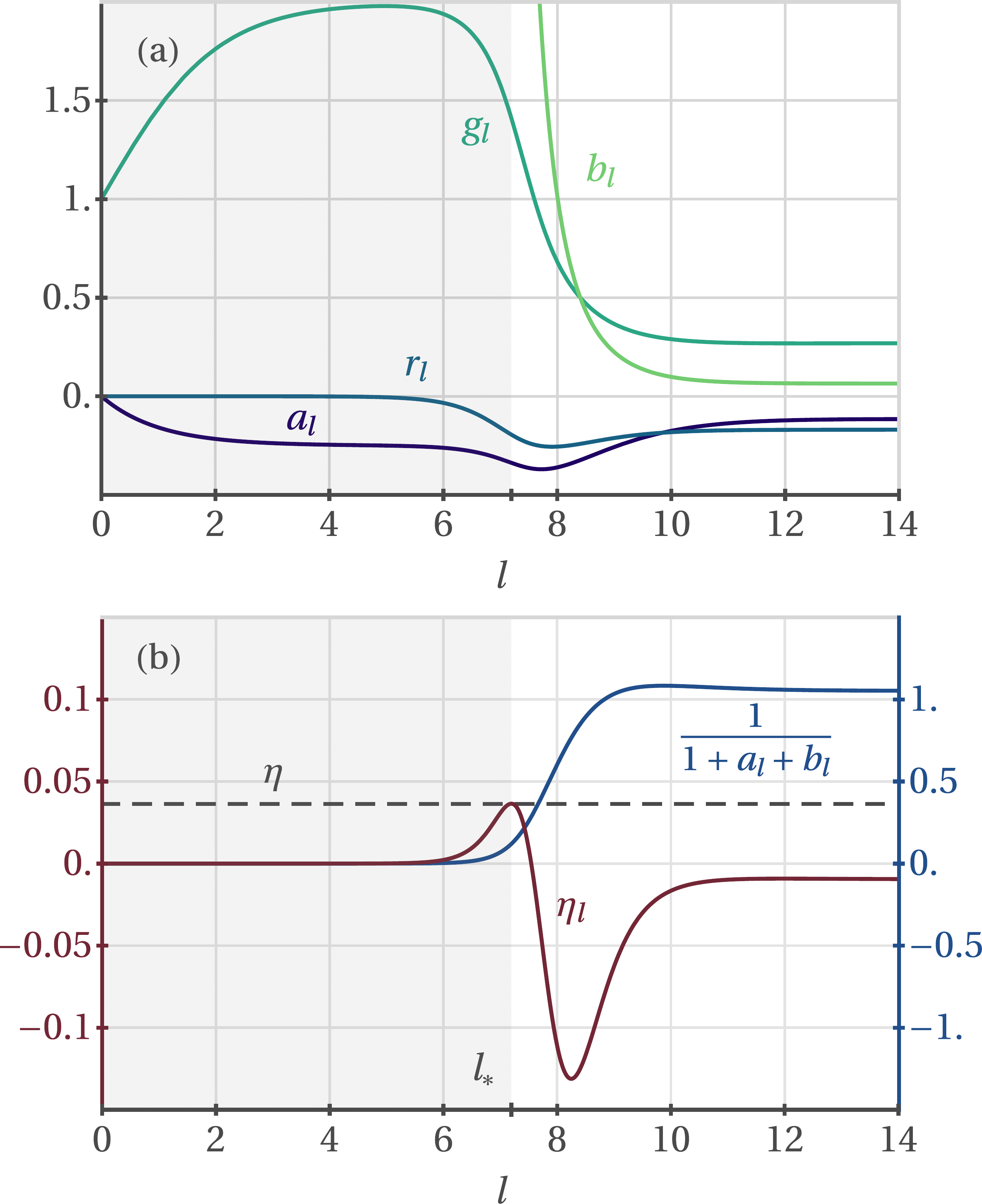}
   \end{center}
  \caption{
(a)
Flow of the couplings $r_l$, $g_l$, $a_l$, and $b_l$ obtained from
Eqs.~(\ref{eq:rlflow})--(\ref{eq:etal}) 
for $D=3$ with initial conditions
$g_0 = 1$, $a_0 =0$, $b_0 =10^{10}$. The initial condition $r_0 = -2.46561 \times 10^{-7} $ is chosen such that the couplings flow into the Wilson-Fisher fixed point $( r_\infty,g_\infty,a_\infty, b_\infty) $.
(b) Flow of the  anomalous dimension  $\eta_l$ given  in Eq.~(\ref{eq:etal})
and control parameter  $ 1/ s_l = 1/(1 +a_l + b_l )$ for $D=3$ and
the same initial conditions as in (a). At $l_{\ast} =7.19416$, where the parameter $1/s_{\ast} =0.11994$ controlling the applied truncation scheme is still small, the scale-dependent $\eta_l$ assumes
a global maximum $\eta_{\ast} =0.03648$.
The dashed line marks the accepted result $\eta=0.03630$ for the three-dimensional Ising model \cite{Kos16}.
}
\label{fig:flow}
\end{figure}

\vspace*{-.35cm}
\section{RG Flow Analysis}
\vspace*{-.3cm}
First of all, we note that the above system of flow equations has a fixed point with one relevant direction and finite $\eta$ which we identify with the Wilson-Fisher fixed point.
In three dimensions,
the numerical values of our rescaled couplings at the fixed point are
 $r_{\infty} =  -0.170$,
 $g_{\infty}  = 0.269$,
 $a_{\infty}  = -0.115$,
 $b_{\infty}  =  0.065$, and
 $\eta_{\infty}  = -0.00957$.
At first sight, it seems that our truncation is not satisfactory, 
as it yields a negative anomalous dimension at the fixed point.
Moreover, keeping in mind that the dimensionless coupling $b_l$ can be identified with the square of the range of the interaction in units of the ultraviolet cutoff, 
we see that at the fixed point the interaction is short range, 
corresponding to  $b_{\infty} \ll 1$. 
It is therefore not surprising that our truncation strategy, 
which relies on the long-range nature of the interaction, 
breaks down as soon as the flowing coupling $b_l$ ceases to be large compared with unity.
On the other hand, for $b_l \gg 1$ our truncation is controlled by the small parameter $1/b_l$ and is expected to be quantitatively precise in this regime.
In fact, by perturbatively calculating the modification of the flow equations (\ref{eq:rlflow})--(\ref{eq:etal}) due to the higher-order vertices shown in Fig.~\ref{fig:flowself} and the momentum-dependent part of the three-point vertex 
$\Gamma^{\psi \varphi \varphi}_{\Lambda} ( \bd{k}_1 ;
 \bd{k}_2 , \bd{k}_3 )$,
we find that all corrections involve at least an additional factor of
$1/s_l = 1/(1 + a_l + b_l )$, 
which is small if  $b_l \gg 1$.
It is then natural to stop the RG flow at some finite scale $l_{\ast}$ where $1/s_{{\ast}}$ is still reasonably small.
If the Ginzburg scale \cite{Ginzburg60,Amit74,Hasselmann07,Kopietz10} 
(which can be identified with the upper limit of the momentum range  where the order-parameter correlation function at the critical point scales as $k^{-2 + \eta}$)
is larger than $\Lambda_{0} e^{ - l_{\ast}}$,
the RG trajectory at $l = l_{\ast}$  already ``feels'' the Wilson-Fisher fixed point so that we expect that the flowing $\eta_l$ at scale $l = l_{\ast}$ will be a reasonable approximation for the critical exponent~$\eta$.

\begin{figure}[t]
\vspace*{-.7cm}
 \begin{center}
  \centering
\vspace{7mm}
 \includegraphics[width=0.435\textwidth]{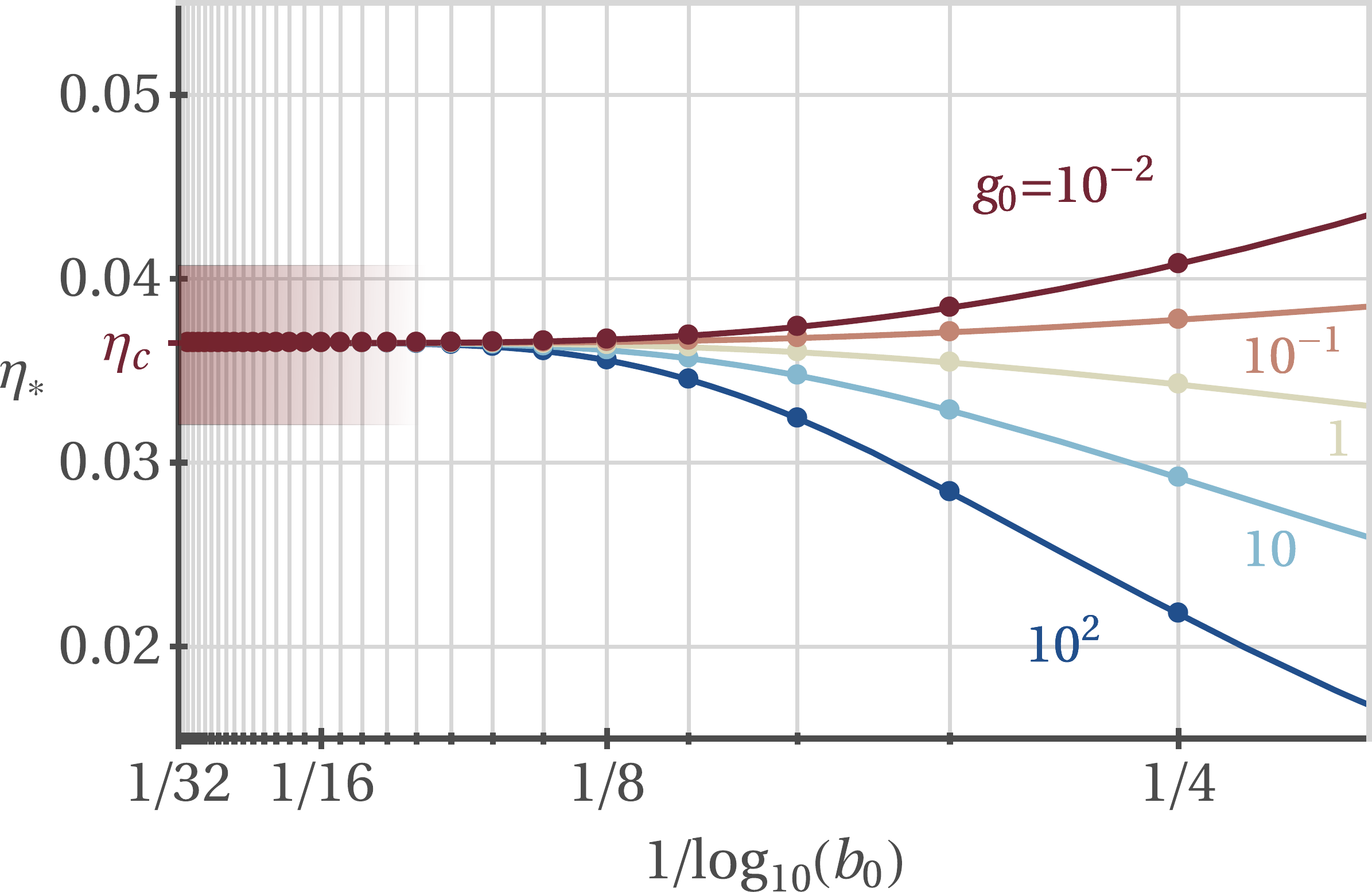}
   \end{center}
\vspace*{-.3cm}
  \caption{
Maximum $\eta_{\ast}$ of $\eta_l$ in $D=3$ for different bare interactions $g_0$ as a function of the bare value $b_0$ of the dimensionless interaction-range parameter $b_l$.
For $1/b_0 \rightarrow 0$ our results for $\eta_{\ast}$ converge to $\eta_c=0.03651$. 
On the scale of the plot this cannot be distinguished from the results of Ref.~\cite{Kos16,Ferrenberg18}.
The shaded region indicates the expected uncertainty $\Delta \eta_c=\eta_cs_c$.}
\label{fig:etaresult}
\end{figure}

To  investigate whether such a scale $l_{\ast}$ really exists, we solve the flow  equations (\ref{eq:rlflow})--(\ref{eq:etal}) numerically.
For $a_0=0$ and given initial values $g_0$ and $b_0$, the initial $r_0$ is thereby  fine-tuned such that for $l \rightarrow \infty$ the RG trajectory flows into the fixed point. 
The flow of the couplings $r_l$, $g_l$, $a_l$, $b_l$, $s_l$, and $\eta_l$
with initial condition $g_0 =1$, and $b_0 =10^{10}$ in $D=3$ is shown in Fig.~\ref{fig:flow}.
The crucial observation is now that the flowing $\eta_l$ exhibits a local maximum $\eta_{\ast}$ at a finite scale $l=l_{\ast}$. 
At this scale the dimensionless parameter $1/ s_{\ast}$ 
which controls the precision of our truncation is still small but rapidly approaches  a number of order unity for $l \gtrsim l_{\ast}$. 
Moreover, from Fig.~\ref{fig:flow} we see that close to the scale $l_{\ast}$ the couplings $r_l$, $g_l$, and $a_l$ exhibit a local extremum before monotonously approaching their fixed point values. 
It is therefore reasonable to assume that the scale $\Lambda_0 e^{ - l_{\ast}}$ defines the boundary of the Ginzburg regime and estimate $\eta \approx \eta_{\ast}$.
As the control parameter $1/ s_{\ast}$ is roughly $0.1$ at this point, we expect that in this way we can obtain $\eta$ with an uncertainty $\Delta\eta/\eta$ of about 
$10 \%$.
We checked that these features are robust with respect to variations of the initial conditions by changing the initial range of the interaction, parametrized by $b_0$, for fixed bare interaction $g_0$. 
While the resulting $l_{\ast}$ grows with increasing $b_0$, we observe a convergence of the corresponding maximum $\eta_{\ast}$.
In Fig.~\ref{fig:etaresult} we present our results for $\eta_\ast$ as a function of the initial value $b_0$ for $g_0$ in the range between $10^{-2}$ and $10^2$. 
The value of $\eta_{\ast}$ obtained in this way converges for $b_0 \rightarrow \infty$ to $\eta_{c} = 0.03651$. 
Amazingly, this is only $1 \% $ larger than currently most precise results \cite{Kos16,Ferrenberg18}, although \textit{a priori} we would have expected agreement only at the  $10 \%$ level. 
While we cannot exclude the possibility that this agreement is accidental, 
we believe that it is caused by a cancellation of the corrections of order in $1/s_l^2$ to the FRG flow equations in $D=3$. 
This point  certainly deserves further attention.

Given the fact that our flow equations (\ref{eq:rlflow})--(\ref{eq:etal}) are valid for arbitrary $D$, we may also use our method to calculate $\eta$ as a function of the dimensionality $D$ of the system. 
\begin{figure}[t!]
\vspace*{-.7cm}
 \begin{center}
  \centering
\vspace{7mm}
 \includegraphics[width=0.5\textwidth]{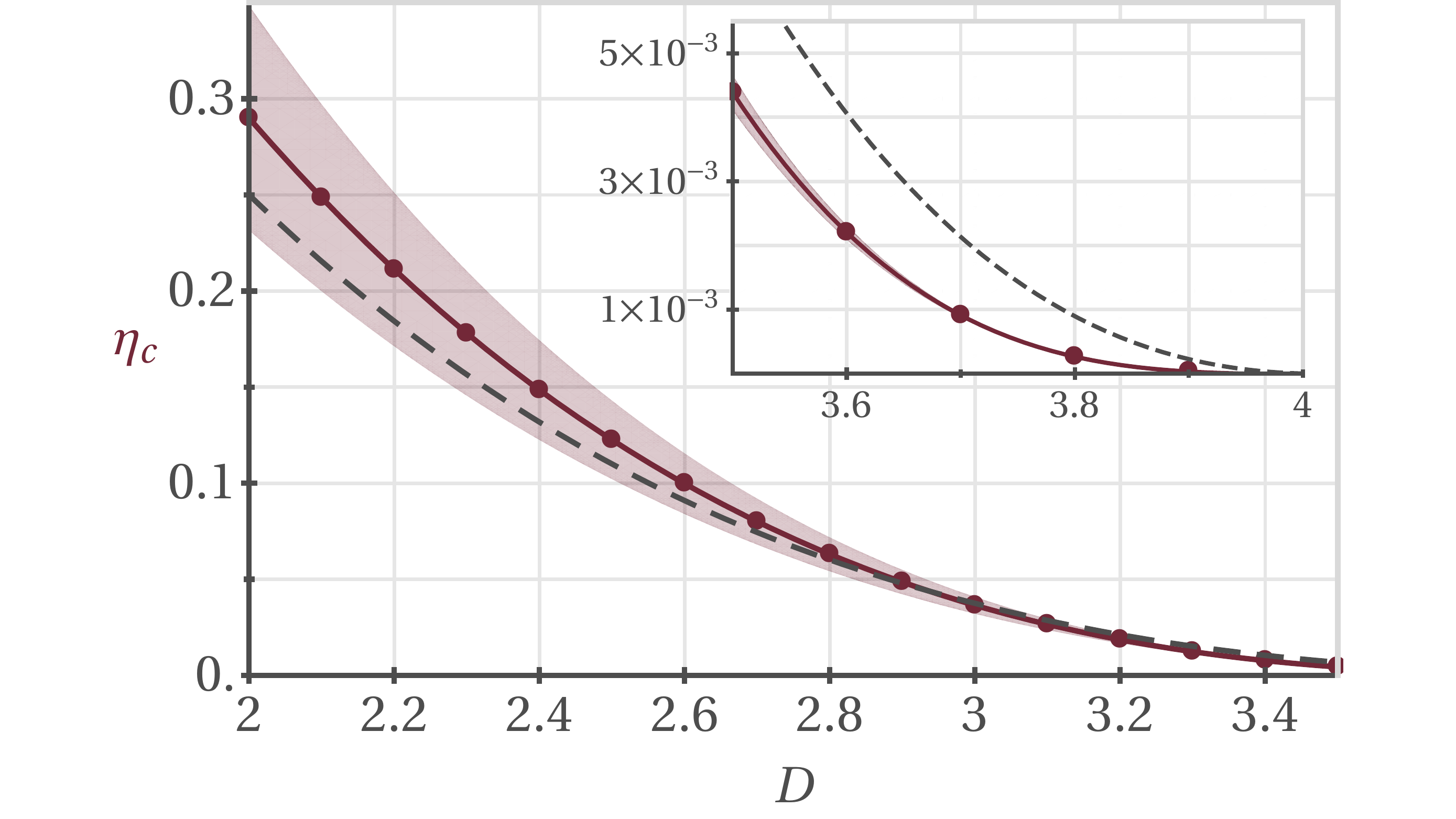}
    \end{center}
\vspace*{-.3cm}
  \caption{
Estimate for the critical exponent $\eta_c(D)$ obtained with our method for dimensions $2 \leq D < 4$, where the shaded region is indicating the expected uncertainty $  \Delta\eta_c(D)=\eta_c(D) s_c(D)$.
The dashed line represents the coinciding result obtained independently by Borel-resumming the $\epsilon$-expansion series \cite{LeGuillou87} and the conformal bootstrap calculations \cite{El-Showk14}.}
\label{fig:etad}
\end{figure}
The result is shown in Fig.~\ref{fig:etad} together with a coinciding result obtained independently via Borel-resumming the $\epsilon$-expansion series \cite{LeGuillou87} and conformal bootstrap calculations \cite{El-Showk14}.
For $D\leq 3$ our value for $\eta_D=\eta_c(D)$ agrees within the expected uncertainty of about $s_D^{-1}$ with the previous results, although the exact value is always slightly larger.
In the opposite limit of small $\epsilon=4-D$ our method is only able to predict the order of magnitude of $\eta_D$.
A natural explanation for this lack of quantitative accuracy for $\epsilon \ll 1$
is that in this case the Ginzburg momentum is exponentially small,
$ k_G \propto e^{ - {\rm{const}} / \epsilon}$
(see Refs.~[\onlinecite{Ledowski04,Kopietz10}]),
so that  the scale where  the renormalized interaction of our model crosses over from long-range to short-range does not overlap with the Ginzburg regime.

\vspace*{-.35cm}
\section{Conclusion}
\vspace*{-.3cm}
In summary, we have developed a method for calculating the critical exponent $\eta$ of the Ising universality class which uses the inverse range of the interaction of an effective Ginzburg-Landau-Wilson model as a small parameter to control the truncation of the vertex expansion of the FRG flow equations.
Although the effective interaction becomes short-range as the RG is iterated,
by stopping the RG flow at a finite scale where the range of the interaction is still large we were able to obtain a surprisingly accurate estimate of $\eta$ in three dimensions (with respect to Refs. \cite{Ferrenberg18,Kos16}).
Note that a similar scheme to obtain the value of the critical exponent $\eta$ was already employed in the context of the $O(2)$ model in two dimensions \cite{Graeter95,Defenu17}.
These works evaluate the corresponding anomalous dimension along a line of unstable pseudofixed points and find a local maximum of $\eta^{O(2)}_{D=2}=0.24$ close to the known value $0.25$. 
The location of the maximum thereby coincides with a crossover from the ordered into the disordered phase, 
similar to our calculation where the extremum of $\eta$ is located at the scale where the interaction changes from long-range to short-range.

An implicit assumption underlying our method is that the Ginzburg regime extends to the scale where the RG flow is stopped to estimate $\eta$.
This assumption seems to be valid in $2 \leq  D \leq 3$, 
but does not hold for small  $\epsilon = 4-D$ where the Ginzburg scale is exponentially small.
Our calculation can be systematically improved by taking the momentum-dependence of the three-point vertex and of  higher order vertex corrections encoded in the different types of induced four-point vertices shown in Fig.~\ref{fig:flowself} into account, 
which gives rise to additional terms in the RG flow equations (\ref{eq:rlflow})--(\ref{eq:etal}) involving higher powers of  the small parameter $1/s_l$. 
If we do not rely on the long wavelength approximation in Eqs. (\ref{eq:sigmalong}) and (\ref{eq:pilong})
our method can be used to calculate the complete momentum-dependence of the self-energy and of the effective interaction.
Finally let us emphasize that the appeal of this work does not only lie in the final precision of the results, which on its own is not impressive compared to sophisticated state-of-the-art methods.
The presented approach is rather an instructive example that it can be possible to extract
critical properties from the Ginzburg regime of the RG flow (and not from the linear regime close to the critical point), thereby allowing for new approximation schemes.
Since the method should furthermore be generally applicable to $O(N)$ models, it may be regarded as a complementary possibility to obtain reasonable estimates for the corresponding $\eta$ with relatively little effort.

\vspace*{-.35cm}
\section*{Acknowledgements}
\vspace*{-.3cm}
We acknowledge financial supportby the Deutsche
Forschungsgemeinschaft (DFG) through SFB/TRR 49 and the hospitality
of the Department of Physics and Astronomy of the University of California, Irvine, where most of this work was accomplished.
%
%

\end{document}